\documentclass{article}

\usepackage{arxiv}
\usepackage{tikz}

\usepackage[utf8]{inputenc} % allow utf-8 input
\usepackage[T1]{fontenc}    % use 8-bit T1 fonts
\usepackage{hyperref}       % hyperlinks
\usepackage{url}            % simple URL typesetting
\usepackage{booktabs}       % professional-quality tables
\usepackage{amsfonts}       % blackboard math symbols
\usepackage{nicefrac}       % compact symbols for 1/2, etc.
\usepackage{microtype}      % microtypography
\usepackage{lipsum}
\usepackage{listings}
\usepackage{graphicx}
\usepackage{amsmath}
\usepackage{bm}

\title{%The Ising zoo
A Performance study of the 2D Ising model on GPUs}

\author{
  \makebox[.47\linewidth]{Joshua Romero} \\
  NVIDIA Corporation\\
  Santa Clara, CA 95050 \\
  \texttt{joshr@nvidia.com} \\
  \And
  \makebox[.47\linewidth]{Mauro Bisson}\\
  NVIDIA Corporation\\
  Santa Clara, CA 95050 \\
  \texttt{maurob@nvidia.com} \\
  \AND
  \makebox[.47\linewidth]{Massimiliano Fatica}\\
  NVIDIA Corporation\\
  Santa Clara, CA 95050 \\
  \texttt{mfatica@nvidia.com} \\
  \And
  \makebox[.47\linewidth]{Massimo Bernaschi}\\
  Istituto per le Applicazioni del Calcolo\\
  National Research Council of Italy\\
  Rome, Italy 00185 \\
  \texttt{massimo.bernaschi@cnr.it} \\
}

\begin{document}
\maketitle

\begin{abstract}
The simulation of the two-dimensional Ising model is used as a benchmark to show the computational capabilities of Graphic Processing Units (GPUs).
The rich programming environment now available on GPUs and flexible hardware capabilities
allowed us to quickly experiment with several implementation ideas: a simple stencil-based algorithm, recasting the stencil operations into matrix multiplies to take advantage of Tensor Cores available on NVIDIA GPUs, and a highly optimized multi-spin coding approach. Using the managed memory API available in CUDA allows for simple and efficient distribution of these implementations across a multi-GPU NVIDIA DGX-2 server. We show that even a basic GPU implementation can outperform current results published on TPUs \cite{TPU2019} and that the optimized multi-GPU implementation can simulate very large lattices faster than custom FPGA solutions \cite{FPGA}.
\end{abstract}

% keywords can be removed

\section{Introduction\label{sec:intro}}
In the past ten years, GPUs have evolved from chips specialized for graphics processing to powerful and flexible accelerators for general computational and data processing tasks. Together with improvements in hardware capabilities, there
has been a growing software ecosystem. In terms of programmability, beyond the original CUDA C, GPUs can now be programmed with compiler directives (OpenACC, OpenMP), or in high level languages such as Python, MATLAB, or Julia. On the system side, there are now servers like the NVIDIA DGX-2 with all GPUs connected via NVLink and NVSwitch. These types of systems are bringing SMP-like capabilities to multi-GPU programming, as GPUs on the node can access data on other GPUs in a fast and transparent way.

In this paper, we use the 2D Ising model to compare the level of performance achievable with different programming approaches on GPUs. %The 2D Ising model has the nice properties of having a theoretical solution, a simple but not trivial access pattern and logic  and the need to generate large set of random numbers.

\section{Ising Model}
In statistical mechanics, ``spin system'' indicates a broad class of models
used for the description of a number of physical phenomena.  A spin system is
usually identified by a Hamiltonian function having the following general form:
\begin{equation}
H = -\sum_{i \ne j}J_{ij}\sigma_i\sigma_j
\label{equ:spinequ}
\end{equation}
The spins $\sigma$ are defined on a {\em lattice} which may have one, two, three
or even a higher number of dimensions. The sum in Equation
\ref{equ:spinequ} runs on the nearest neighbors of each spin location (for example, 2
in 1D, 4 in 2D and 6 in 3D). The spins and the couplings $J$ may be either discrete or continuous and their values determine the
specific model. In the Ising model \cite{Ising} of ferromagnetism, the spins can be in one of two states (+1 or -1), $J_{ij}$ is $>0$ and, as a further simplification, all of the nearest neighbors $<ij>$ have the same interaction strength so that $J_{ij} = J$ for all pairs $i, j$. There are analytical solutions for the Ising model in 1D and 2D; however, these can be considered an exception because, despite the illusory simplicity of their Hamiltonian formulation, the study of spin systems in higher dimensions is by no means trivial. Most of the time, numerical simulations (very often based on
Monte Carlo methods) are the only way to understand their behavior.

A Monte Carlo simulation of the Ising model can be performed with the Metropolis algorithm \cite{Metropolis}, where
the following steps are repeated:
\begin{itemize}
    \item From an initial configuration of spins, flip a randomly chosen spin.
    \item If the change in energy between the old and new state is negative, we accept the change.
    \item Otherwise, the move is accepted with probability $e^{(-\beta \Delta E)}$, where $\beta$ is the inverse of the temperature of the system and $\Delta E$ is the difference in the energy due to the spin flip.
\end{itemize}

It is very desirable from a computational point of view to update multiple spins in parallel.
Given the local interactions of a spin with its neighbors, we can see that if we consider the lattice as a checkerboard,
the flipping of a spin of a particular color is completely determined by its neighbors of the opposite color.
We can update all the spins of one color in parallel, keeping the other color constant, and then repeat the process with the opposite color.

The checkerboard decomposition can be used to run parallel versions of other \textit{local} Monte Carlo algorithms, like the \textit{Heat Bath} algorithm in which the probability $P$ of a spin flip from $\sigma$ to $-\sigma$ is equal to $\nicefrac{e^{-\beta \Delta E}}{e^{-\beta \Delta E}+1}$. Due to their locality, algorithms like Metropolis and Heat Bath suffer from the so called \textit{critical slowing down} syndrome; a state in which it becomes increasingly difficult to flip a spin at random as it is likely to be coupled
to neighboring spins pointing in the same direction.
A solution to this problem is provided by an algorithm proposed by U. Wolff \cite{Wolff1989} in which a whole cluster of spins is flipped in each update instead of a single one.
The cluster is constructed starting from a seed spin selected at random and looking at its neighboring spins. Those with the same sign as the seed
spin are added to the cluster with a probability $P_{add}$ equal to $1-e^{-2\beta J}$, whereas they are excluded from
the cluster with probability $1 - P_{add}$  (spins having the opposite value with respect to the seed spin are ignored). This implies that spins are added
to the cluster with a probability that is temperature dependent. It is not difficult to check that in both the low and high
temperature regimes, the Wolff algorithm is not very efficient and the simpler Metropolis algorithm performs better. As a consequence, there is still much interest in using efficient (from the computational viewpoint) implementations of the Metropolis algorithm for the simulation of the Ising (and similar) models \cite{Bernaschi2012}. 

In the present paper, we discuss the performance of several implementations for the simulation of the 2D Ising model running on NVIDIA Volta GPUs. We focus on the 2D model because  the results can be easily compared to the analytical solution derived by Onsager \cite{Onsager}. We compare our performance with those recently produced on other computing platforms \cite{TPU2019,FPGA}.

\section{Single-GPU implementations}
\subsection{Basic Implementation}
To begin, a basic single-GPU implementation of the checkerboard Metropolis algorithm was implemented in Python, combining features available in the popular \texttt{Numba} and \texttt{CuPy} packages. Specifically, \texttt{Numba} \cite{Numba} was used for general GPU data handling via its provided \texttt{device\_array} objects and also for its ability to compile custom CUDA kernels at runtime expressed purely in Python. In our experimentation, we found that the random number generation supported in \texttt{Numba} for use in device kernels to be fairly low performing. As a replacement, we used available bindings into the NVIDIA \texttt{cuRAND} \cite{cuRAND} library available in \texttt{CuPy} \cite{CuPy} to pre-populate an array of random numbers as a separate operation before each lattice update kernel call.     

\begin{figure*}[t]
\begin{center}
        \includegraphics[scale=0.75]{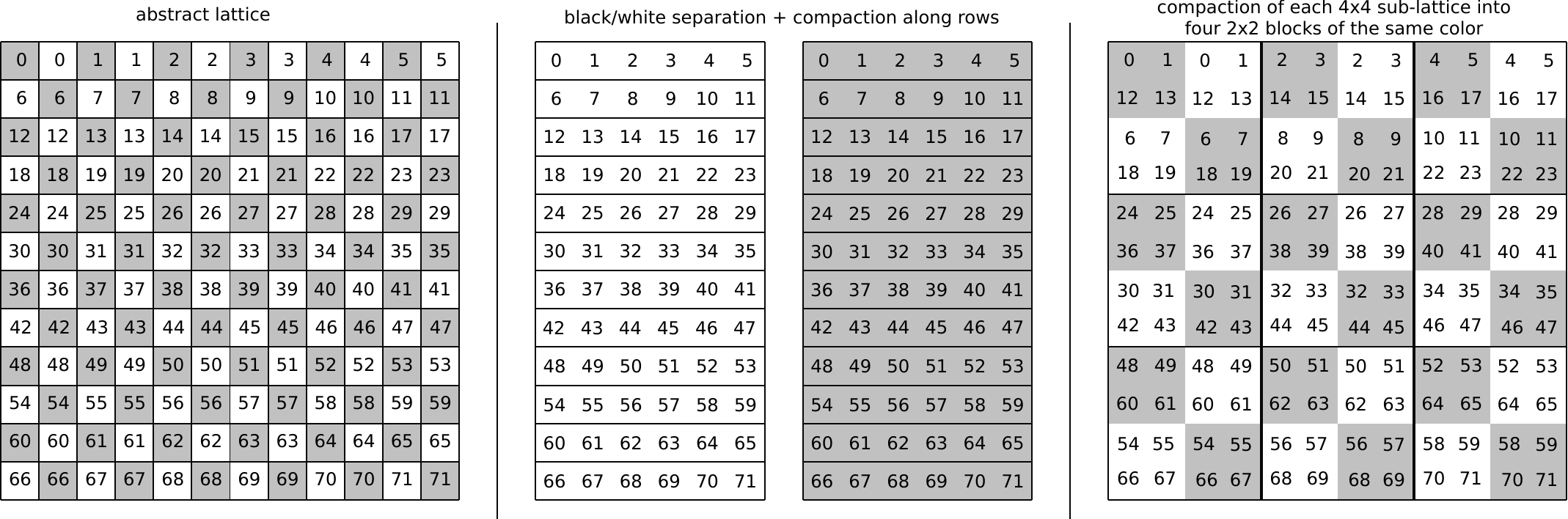}
\end{center}
        \caption{On the left, an abstract $12\times 12$ lattice with the checkerboard pattern highlighted is shown. In the center, the lattice is represented using two arrays, each containing one color of spins compacted along the rows. On the right, the lattice is decomposed into $4\times 4$ sub-lattices, where each sub-lattice is represented as a sequence of four $2\times 2$ blocks of spins of the same color.}
        \label{fig:checker}
\end{figure*}

The $N \times N$ checkerboard lattice of spins is represented in two separate arrays of dimension $N \times N/2$, with each array containing only spins of a single color. This decomposition is depicted in the central image in Figure \ref{fig:checker}. As each spin only takes the value of $\pm 1$, each spin location can be represented using a single byte. While further compressed data representations are possible, a byte is the smallest data type that does not require bitwise operations.

With the data decomposition in place, the implementation is straightforward and consists of two steps per color, per iteration. For a given color:
\begin{enumerate}
    \item Populate an $N \times N/2$ array of random values with \texttt{CuPy/cuRAND}
    \item Update spins on the lattice for the current color using the opposite colored lattice spin values and the random value array, implemented in a custom kernel written with \texttt{Numba}. See Figure \ref{fig:update_lattice_listing} for a listing of the spin update code. 
\end{enumerate}

To better gauge the performance of \texttt{Numba}, a nearly identical implementation of this basic algorithm for single-GPU was also implemented in CUDA C. A comparison of the main lattice kernel in both implementations can be seen in Figure \ref{fig:update_lattice_listing}. 

\begin{figure}
\begin{minipage}{0.49\textwidth}
\centering
\begin{lstlisting}[language=Python]
@cuda.jit
def update_lattice(lattice, op_lattice, randvals, is_black):
  n,m = lattice.shape
  tid = cuda.blockIdx.x * cuda.blockDim.x + cuda.threadIdx.x
  i = tid // m
  j = tid % m

  if (i >= n or j >= m): return

  # Set stencil indices with periodicity
  ipp = (i + 1) if (i + 1) < n else 0
  jpp = (j + 1) if (j + 1) < m else 0
  inn = (i - 1) if (i - 1) >= 0 else (n - 1)
  jnn = (j - 1) if (j - 1) >= 0 else (m - 1)

  # Select off-column index based on color and row index parity
  if (is_black):
    joff = jpp if (i % 2) else jnn
  else:
    joff = jnn if (i % 2) else jpp

  # Compute sum of nearest neighbor spins
  nn_sum = op_lattice[inn, j] + op_lattice[i, j] + op_lattice[ipp, j] + op_lattice[i, joff]

  # Determine whether to flip spin
  lij = lattice[i, j]
  acceptance_ratio = math.exp(-2.0 * inv_temp * nn_sum * lij)
  if (randvals[i, j] < acceptance_ratio):
    lattice[i, j] = -lij

\end{lstlisting}
\end{minipage}
\begin{minipage}{0.49\textwidth}
\centering
\begin{lstlisting}[language=c]
template<bool is_black>
__global__ 
void update_lattice(char* lattice,
                    const char* __restrict__ op_lattice,
                    const float* __restrict__ randvals,
                    const float inv_temp,
                    const int nx,
                    const int ny) {
  const int tid = blockDim.x * blockIdx.x + threadIdx.x;
  const int i = tid / ny;
  const int j = tid % ny;

  if (i >= nx || j >= ny) return;

  // Set stencil indices with periodicity
  int ipp = (i + 1 < nx) ? i + 1 : 0;
  int inn = (i - 1 >= 0) ? i - 1: nx - 1;
  int jpp = (j + 1 < ny) ? j + 1 : 0;
  int jnn = (j - 1 >= 0) ? j - 1: ny - 1;

  // Select off-column index based on color and row index parity
  int joff;
  if (is_black) {
    joff = (i % 2) ? jpp : jnn;
  } else {
    joff = (i % 2) ? jnn : jpp;
  }

  // Compute sum of nearest neighbor spins
  char nn_sum = op_lattice[inn * ny + j] + op_lattice[i * ny + j] + op_lattice[ipp * ny + j] + op_lattice[i * ny + joff];

  // Determine whether to flip spin
  char lij = lattice[i * ny + j];
  float acceptance_ratio = exp(-2.0f * inv_temp * nn_sum * lij);
  if (randvals[i*ny + j] < acceptance_ratio) {
    lattice[i * ny + j] = -lij;
  }
}

\end{lstlisting}
\end{minipage}
\caption{Single-GPU \texttt{update\_lattice} kernel in Python (left) and CUDA C (right) used in basic implementations. \texttt{@cuda.jit} decorator used to enable \texttt{Numba} JIT compilation of CUDA kernel.}
\label{fig:update_lattice_listing}
\end{figure}

\subsection{Tensor Core Implementation}
In \cite{TPU2019}, an implementation of the checkerboard Metropolis algorithm was developed to map the computation of nearest-neighbor sums to matrix multiplications in order to execute them on TPUs using their hardware FP16 matrix multiply units (TPU Tensor Cores). As GPUs are generally programmable and the computation can be directly expressed in CUDA, such a mapping to matrix multiplies is not necessary for GPUs and adds unneeded complexity; however, as a means for comparison, a version of this matrix multiply algorithm was implemented to utilize NVIDIA Tensor Cores available on Volta GPUs. A complete description of the algorithm can be found in \cite{TPU2019} and only key details will be discussed here. For this implementation, the lattice data is organized in a pattern similar to the right-most diagram in Figure \ref{fig:checker}. In our implementation, the lattice is organized into $256 \times 256$ sub-lattices, each further decomposed into four $128 \times 128$ blocks. At the high level, the algorithm uses matrix multiplications to compute sub-lattice local nearest neighbor sums using a kernel matrix $\bm{K}$ of the form,
\begin{equation}
    \bm{K} = 
    \begin{bmatrix} 
    1 & 1 & 0 & \dots & 0 & 0 & 0 \\
    0 & 1 & 1 & \dots & 0 & 0 & 0 \\
    0 & 0 & 1 & \dots & 0 & 0 & 0 \\
    \vdots & \vdots & \vdots & \ddots & \vdots & \vdots & \vdots \\
    0 & 0 & 0 & \dots & 1 & 0 & 0 \\
    0 & 0 & 0 & \dots & 1 & 1 & 0 \\
    0 & 0 & 0 & \dots & 0 & 1 & 1 \\
    \end{bmatrix}_{128 \times 128}
\end{equation}
For a given sub-lattice, $\bm{\sigma}^{ij}$, to compute the sum of sub-lattice local nearest neighbor spins for the black sub-blocks, $\bm{\sigma}^{ij}_{00}$ and $\bm{\sigma}^{ij}_{11}$, the following operations are performed,
\begin{align}
    nn_\text{L}(\bm{\sigma}^{ij}_{00}) &= \bm{\sigma}^{ij}_{01} \bm{K} + \bm{K}^T \bm{\sigma}^{ij}_{10} \label{eq:b0}\\
    nn_\text{L}(\bm{\sigma}^{ij}_{11}) &= \bm{\sigma}^{ij}_{10} \bm{K}^T + \bm{K} \bm{\sigma}^{ij}_{01} \label{eq:b1}
\end{align}
where $nn_\text{L}()$ denotes the sum of sub-lattice local nearest neighbor spins. A similar expression can be used to compute sums for the white sub-blocks,
\begin{align}
    nn_\text{L}(\bm{\sigma}^{ij}_{10}) &= \bm{\sigma}^{ij}_{11} \bm{K} + \bm{K} \bm{\sigma}^{ij}_{00} \label{eq:w0}\\
    nn_\text{L}(\bm{\sigma}^{ij}_{01}) &= \bm{\sigma}^{ij}_{00} \bm{K}^T + \bm{K}^T \bm{\sigma}^{ij}_{11} \label{eq:w1}
\end{align}
With these equations, one can obtain the sub-lattice local sums for a given color over the entire lattice via two batched matrix-multiply operations, corresponding to the left and right summands in either Eqs. \ref{eq:b0} and \ref{eq:b1} for the black spins, or Eqs. \ref{eq:w0} and \ref{eq:w1} for the white spins. To enable the use of tensor cores, the lattice spins and nearest neighbor sums are stored in half-precision, and the matrix multiplies are computed using \texttt{cublasHgemmBatched} routine available in the NVIDIA \verb|cuBLAS| library \cite{cuBLAS}. 
Once the sub-lattice local sums are obtained for a given color, a separate kernel adding boundary contributions from neighboring sub-lattices is executed, followed by a final spin update kernel which uses the completed sums and generated random values to update spins. In this implementation, we used the \verb|Philox4_32_10| generator provided by the device API of the \verb|cuRAND| library to generate per-thread random numbers within the spin update kernel. In order to avoid allocated separate global memory to store generator states and loading/storing them for every color update, we initialize a new state at each kernel call in a way such that each thread progresses coherently along the same random sequence across consecutive launches. This is possible thanks to the \verb|curand_init()| call that allows to specify a \verb|seed|, a \verb|sequence|, and an \verb|offset|. The seed determines a series of random sequences, the sequence number identifies one of those sequences, and the offset specifies an element in that sequence. For each kernel call, each thread uses the same seed, specifies as sequence number its unique linear index in the grid, and specifies an offset equal to the total count of random numbers generated in the previous kernel calls. This is in contrast to the approach in the basic implementation where an array of random values is pre-populated using the host API. To summarize, the steps to update spins for a given color are as follows:
\begin{enumerate}
    \item Compute sub-lattice local spin sums for current color using two batched matrix-multiplication operations (via calls to \texttt{cublasHgemmBatched})
    \item Add boundary contributions from neighboring sub-lattice to local sums using a custom boundary update CUDA kernel. 
    \item Update spins for current color using completed sums and random values generated with \verb|cuRAND| in a custom spin update CUDA kernel. 
\end{enumerate}

From this description, a few notable issues with this approach can be discussed. First, the use of matrix multiplications necessitates additional memory usage in order to store intermediate local nearest neighbor sums (for at least half of the total lattice). On top of this, to utilize tensor cores the sums and lattice spins must be stored in half-precision which further increases memory requirements. Next, the splitting of the computation into several distinct operations results in increased global device memory traffic as data must be loaded and stored from global memory in between operations. Additionally, the standalone boundary update operation yields many uncoalesced memory loads and stores when updating the sub-lattice boundaries along the direction strided in memory, resulting in poor memory bandwidth utilization for this kernel. As this problem is memory-bandwidth limited, increasing memory bandwidth pressure just to enable the mapping of the problem to tensor cores is not a wise trade-off. This is especially true when the matrix multiplications themselves consist of mostly useless FLOPs (i.e. multiplications by zero), with only two multiplications per inner product contributing to the final result, yielding a fraction of $1/64$ useful FLOPs when using $128 \times 128$ matrix multiplications.
%Regardless, it is interesting to note that this implementation has higher performance in term of flops with respect to  the classical stencil approach, but most of the flops  are "wasted" (with the simple stencil of the 2D Ising model, the coefficient matrix is very sparse and a different matrix multiply is needed for each directions).\\
%SOME WORDS ON THE BLOCKED REPRESENTATION OF THE LATTICE AND REFERENCE TO FIG. 1?

\subsection{Optimized Implementation}

\begin{figure*}[t]
\begin{center}
        \includegraphics[scale=0.92]{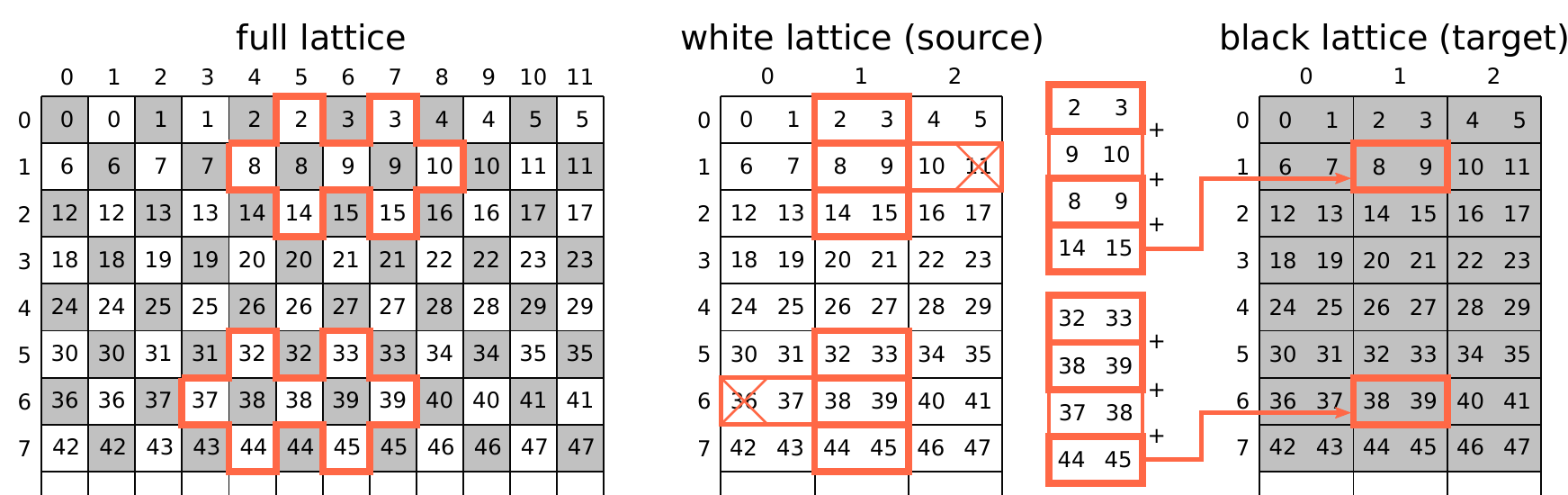}
\end{center}
        \caption{Example of the memory accesses required to load the neighboring spins in the multi-spin case. The left side contains the abstract lattice of spins, colored according to the checkerboard pattern. Represented to the right are the actual arrays of white and black spins as they are stored in memory. Each color is compacted along the rows and stored using two spins per word. There are two possible access patterns required to load neighboring spins based on the color of target spins and on the parity of the row index of the word containing them. Assuming the target lattice is the black one, let's consider an odd row case, for example $(8, 9)_B$, and an even row case, for example $(38, 39)_B$. In both cases, all the top and bottom spins are found in the two words below and above the target word ($(2,3)_W$, $(14,15)_W$ for $(8,9)_B$, and $(32,33)_W$, $(44,45)_W$ for $(38,39)_B$). Of the neighboring spins lying on the same row, all but one are always located in the word at the same coordinates as target word ($(8,9)_W$ and $(38,39)_W$ white). The last spin is either the leftmost one in the word to the right of the central one, if the row index is odd ($(10,xx)_W$ for $(8,9)_B$), or the rightmost one in the word to the left, if the row index is even ($(xx,37)_W$ for $(38,39)_B$). It is easy to see that these cases for the side words are reversed when the target lattice is the white one.}
        \label{fig:multispin_layout}
\end{figure*}

In order to use the compute resources of the GPU as efficiently as possible, we developed an optimized implementation of the Metropolis algorithm based on the multi-spin coding technique \cite{Multispin} in CUDA C. Similar to the basic implementation, we represented the spin lattice of size $N\times M$ as two separate arrays of size $N\times M/2$, each holding one set of spins from the checkerboard representation. Each spin is represented with $4$ bits instead of using an entire byte (or larger words). This choice not only reduces the memory footprint of the lattice, with respect to using a whole word per spin, but it makes it possible to perform the sums of neighbors' values for multiple consecutive spins (of the same color) in parallel, provided that the theoretical spin values -1/1 are mapped to 0/1, respectively. The combination of the reduced number of bits used to store a spin value with the use of the largest word size (64-bit) allows drastic reductions in the number of addition instructions executed by the hardware. Assuming that each of four 64-bit words contain 16 consecutive spins from the N, S, E, and W directions, three additions are sufficient to compute the neighbors sums instead of the $48$ that would be required if each quadruple was processed independently ($16\times 3$). The choice of four bits per spin enables performing the sums without using additional variables given that each spin can contribute at most $1$. In theory, since each sum can add to at most $4$, three bits would be sufficient but that would require explicit handling of the inter-word cases with no sizeable benefit as far as performance or storage are concerned.

The update of each color is performed with a single kernel that is launched with enough blocks to cover the whole lattice. Each thread is in charge of updating $64$ spins (256 bits), thus it executes two 128-bit loads from the target lattice array (the spins to be updated), eight additional 128-bit loads from the source lattice array (to fetch the four neighbors of the target spins, four loads per source word), and two 128-bit writes to store the flipped spins back into the target lattice. The separation on the spins according to the checkerboard pattern is such that it is quite simple to compute the words required to load the neighbors of a target word of spins.
The neighbors of the spins in word $(i,j)$ from the target lattice are found in the words from the source lattice at coordinates $(i-1,j)$, $(i,j)$, $(i+1,j)$ (three vertically aligned words centered on $(i,j)$) and in a word from one of the sides. If $i$ is even, then this word is the one at coordinates $(i,j-1)$, otherwise it's at $(i,j+1)$. The side word contains only one spin required for the computation (the one nearest to the central word) and so it cannot be directly used in the neighbors additions. This however, can easily be fixed by shifting out the unnecessary spins and shifting in an equal number of spins from the central word, as shown in Figure \ref{fig:multispin_layout}. 

In order to avoid reading the lattices entries multiple times from the global memory, each block initially loads its spin tile from the source lattice into the shared memory, and then the threads load the words containing their neighboring spins from that faster memory space. We generate random numbers similarly to what we described for the tensor core implementation, using the  \verb|Philox4_32_10| generator provided by the device API of the \verb|cuRAND| library. 
%For what concerns the random source, we used the \verb|Philox4_32_10| generator provided by the device API of the \verb|cuRAND| library \cite{cuRAND}. In order to avoid reserving global memory for the states and loading/storing them for every color update, we initialize a new state at each kernel call in a way such that each thread progresses coherently along the same random sequence across consecutive launches. This is possible thanks to the \verb|curand_init()| call that allows to specify a \verb|seed|, a \verb|sequence|, and an \verb|offset|. The seed determines a series of random sequences, the sequence number identifies one of those sequences, and the offset specifies an element in that sequence. So for each kernel call, each thread use the same seed, specifies as sequence number its unique linear index in the grid, and specifies an offset equal to the total count of random numbers generated in the previous kernel calls.
Each simulation step is performed by performing two kernel launches, one for the black and one for the white sublattice.

\section{Multi-GPU implementations}

A classic multi-GPU approach would implement the update of the halo spins in a different routine from the spins in the bulk to allow an overlap of communication and computation, as shown in \cite{Bernaschi2012}. While this approach is very effective and allows good scaling, 
in this work  we exploited new capabilities available in the DGX line of systems (DGX Station, DGX-1 and DGX-2), where the GPUs are connected through NVLink and unified memory allows allocations that span multiple GPUs. When one GPU needs to access data stored on another GPU, the driver will automatically migrate the corresponding memory pages.

When using multiple GPUs, the whole lattice can be partitioned into horizontal slabs and each GPU stores one slab in its own global memory in the same layout employed in the single-GPU case (according to the checkerboard pattern). In this way, each GPU needs only read access to the memory of the two GPUs that handle the slabs on top and bottom of its own region. 

\subsection{Basic Implementation}
First, we consider distributing the basic implementation to multiple GPUs. One limitation of the \texttt{Numba} library is that a single thread can have at most one active CUDA context at a time, which actually prohibits the simple usage of unified memory to access data across multiple GPUs. Instead, a multi-process multi-GPU version was implemented using MPI via \texttt{mpi4py} and CUDA Interprocess Communication (IPC) handles, obtained directly from the \texttt{Numba} \texttt{device\_array} objects. In this case, each process obtains a CUDA IPC handle to the lattice allocations of neighboring slabs, and uses them to access spin data in the update lattice kernel when the neighbor stencil crosses slab boundaries.    

\subsection {Tensor Core and Optimized Implementations}
Since both these implementations are written in CUDA C, the full set of options for distribution across multiple GPUs are available.  
%Unified memory was the most straightforward option and was used in both implementations to distribute the computation across multiple GPUs in a single compute server. %With the availability of fast peer-to-peer device connections via NVLink and NVSwitch on a DGX-2 server, this approach was found to be very performant. 

\begin{figure*}[t]
\begin{center}
        \includegraphics[scale=0.85]{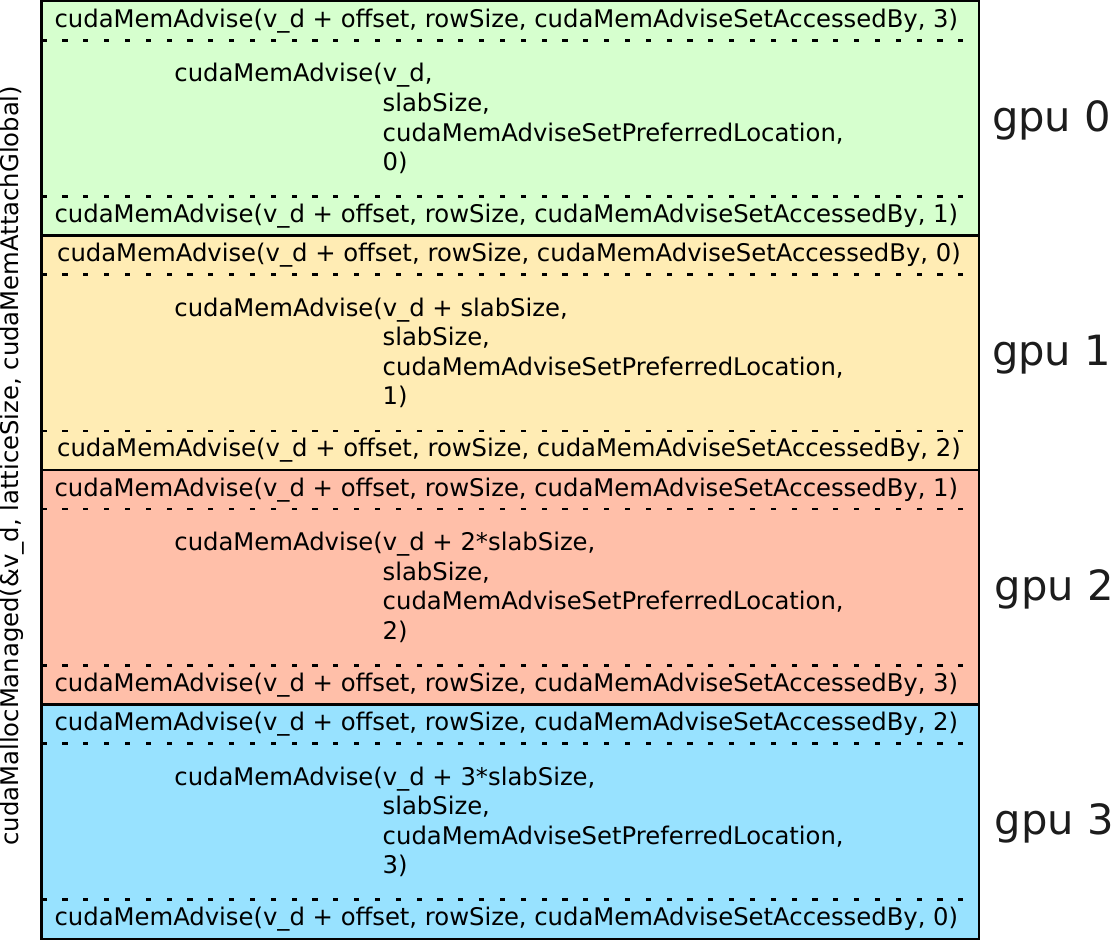}
\end{center}
        \caption{Example of Unified Memory calls required to process a squared lattice with 4 GPUs enabling direct memory access to boundary regions between pairs of neighboring devices via NVLink.}
        \label{fig:multi_gpu_scheme}
\end{figure*}

%For the tensor core implementation, this distribution is transposed into vertical slabs due to the use of a column-major data layout. 
The slab based distribution can be realized very easily by using the CUDA Unified Memory system. It is sufficient to allocate the entire $N\times N$ lattice via a single call to the \verb|cudaMallocManaged()| function and then, for each slab, setting the preferred location to be the memory belonging to target GPU and setting up the mapping of the first and last line of that slab in the page tables of adjacent GPUs. These operation are performed by using the \verb|cudaMemAdvise()| call by specifying, respectively, the \verb|cudaMemAdviseSetPreferredLocation| and the \verb|cudaMemAdviseSetAccessedBy| advice parameters. In total, since black and white lattices are stored in separate arrays, six \verb|cudaMemAdvise()| calls are required per GPU. Figure \ref{fig:multi_gpu_scheme} shows an example of the Unified Memory calls and the lattice regions they need to be called onto.

This setup does not require any change to the existing GPU kernels to access external memory, or any explicit exchange of data among GPUs (like it would normally be necessary in a CUDA+MPI setup). Rather, the existing kernels are simply launched on each GPU with the lattice array accesses properly offset to access different slabs. When a GPU executes a load on a word belonging to one of its neighbors then a data transfer through NVLink automatically takes place.

\section{Results}
\subsection{Single-GPU Performance}

In this section, we present single-GPU performance results. We ran our single-GPU tests on a Tesla V100-SXM card mounted in a DGX-2 \cite{DGX2}. The card is equipped with 32GB of HBM2 memory and a total of $5120$ CUDA cores. For the basic and tensor core implementations, we ran tests on lattice sizes ranging from $(20 \times 128)^2$ to $(640 \times 128)^2$ to compare directly with TPU results reported in \cite{TPU2019}. For the optimized implementation, we ran tests with different lattice sizes ranging from $2048^2$ (2 MB) to $(64\times 2048)^2$ (8 GB), quadrupling the total number of spins at each step. We also ran the largest lattice possible with the available memory, a $(123\times 2048)^2$ requiring 30.3 GB of memory.  The results are reported in Tables \ref{singleGPU_res_python_tc} and \ref{singleGPU_res} respectively.
% From the google paper:
% Dividing flips/ns by number of cores, the flips per nanosecond %per TPUv3 core is roughly % 11.4337, compared to 3.2188 per GPU—a 250% speedup.
% So, let's give them a taste of their one medicine. Hold your belly here...

Considering the performance results in Table \ref{singleGPU_res_python_tc}, both the basic implementation and the tensor core one achieve higher performance on a single Tesla V100-SXM than the single TPUv3 results reported in \cite{TPU2019}, with speedups ranging from a factor of 3 for the tensor core implementation to 5 for the CUDA C variant of the basic implementation, when comparing highest reported spin update rates. Comparing across GPU implementations in this table, the tensor core implementation and basic implementation in Python are significantly slower than the basic implementation written in CUDA C. For the tensor core implementation, this is not surprising due to the additional memory bandwidth overhead introduced from splitting the problem into more distinct operations and poor performance of the boundary update kernel due to the required uncoalesced loads and stores. In comparing the basic implementation in Python to the implementation in CUDA C, the major performance difference was mostly due to quality of the compiled spin update kernels. While the Python and CUDA C versions of these kernels are nearly identical (see Figure \ref{fig:update_lattice_listing}), the JIT-compiled Numba kernels were much less performant, in part due to greater register usage relative to the kernel generated by the \texttt{NVCC} compiler with the CUDA C source.  

For comparison with the performance of our optimized implementation, we include in Table \ref{singleGPU_res} the best results from the high performance TPUv3 implementation on a single and 32 TPUv3 cores, as well as the performance achieved on a FPGA \cite{FPGA} using a lattice of size exactly $1024^2$. The comparison shows that our optimized implementation on a single V100-SXM provides a speedup in excess of $3000\%$ with respect to a TPUv3 core. It is necessary to combine the processing power of 32 TPUv3 cores (four TPU units) to reach performance in the ballpark of a single Tesla V100. 

\begin{table}[ht]
    \centering
    \begin{tabular}{r|cccc}
        lattice size        & Basic (Python) & Basic (CUDA C) & Tensor Core & TPU (on TPUv3 core) \cite{TPU2019}  \\ \hline
        $ (20 \times 128)^2$  & 15.179 & 48.147 & 31.010 & 8.1920 \\
        $ (40 \times 128)^2$  & 40.984 & 59.606 & 35.356 & 9.3623 \\
        $ (80 \times 128)^2$  & 42.887 & 64.578 & 38.726 & 12.336 \\
        $ (160 \times 128)^2$ & 43.594 & 66.382 & 39.152 & 12.827 \\
        $ (320 \times 128)^2$ & 43.768 & 66.787 & 39.208 & 12.906 \\
        $ (640 \times 128)^2$ & 43.535 & 66.954 & 38.749 & 12.878 \\ \hline
    \end{tabular}
    \vspace{.1in}
    \caption{Single-GPU performance comparison between basic, tensor core, and TPU implementations. Results reported in flips per nanosecond on the same lattices used in \cite{TPU2019}.}\label{singleGPU_res_python_tc}
\end{table}

\begin{table}[ht]
    \begin{minipage}{0.30\linewidth}
        \centering
        \begin{tabular}{rr}
                   lattice size        & flip/ns \\ \hline
                  $ (1\times 2048)^2$  &  231.09 \\
                  $ (2\times 2048)^2$  &  318.95 \\
                  $ (4\times 2048)^2$  &  379.27 \\
                  $ (8\times 2048)^2$  &  411.65 \\ 
                  $(16\times 2048)^2$  &  420.44 \\
                  $(32\times 2048)^2$  &  420.77 \\
                  $(64\times 2048)^2$  &  418.23 \\ 
                 $(123\times 2048)^2$  &  417.53 \\\hline\hline
        1 TPUv3 core in \cite{TPU2019} &   12.91 \\
      32 TPUv3 cores in \cite{TPU2019} &  336.01 \\\hline
           FPGA ($1024^2$) \cite{FPGA} &  614.40\footnotemark{}\\
        \end{tabular}
    \end{minipage}
    \hspace{0.1in}
    %\hfill
    \begin{minipage}{0.62\linewidth}
        \centering
        \includegraphics[scale=0.43]{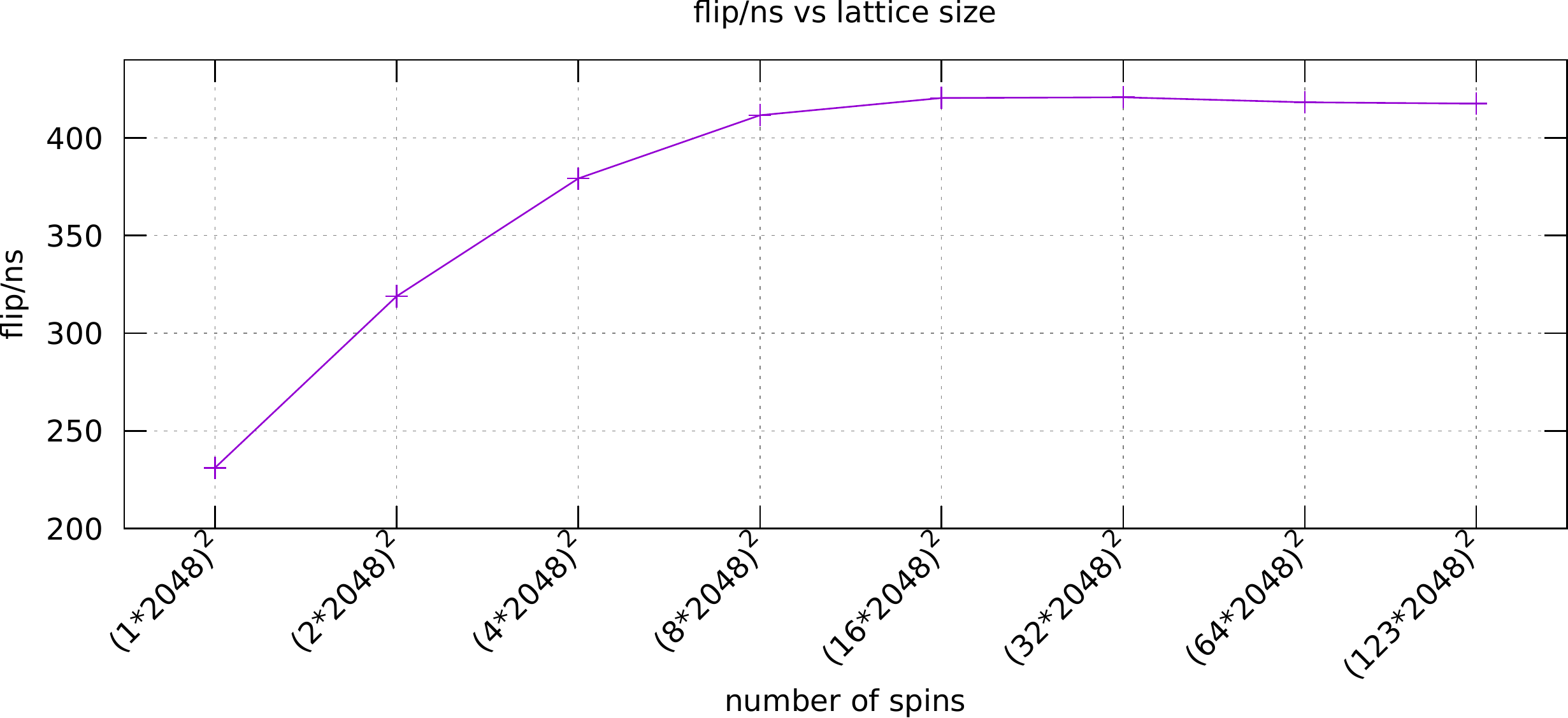}
    \end{minipage}
    \vspace{.1in}
    \caption{Flips per nanosecond obtained by the optimized multi-spin code on a single Tesla V100-SXM card with different lattice sizes, requiring an amount of memory ranging from 2MB to 30GB. For comparison purposes, the table also reports the best timings with 1 and 32 TPUv3 cores from \cite{TPU2019}, and with 1 FPGA from \cite{FPGA}.} \label{singleGPU_res}
\end{table}

\subsection{Multi-GPU Performance}

In this section, we present the multi-GPU performance results. For the optimized code, we measured performance on both a DGX-2 and a DGX-2H system. The DGX-2H is a newer version of the DGX-2 server that is outfitted with higher-performing CPUs and GPUs. Its 16 Tesla V100 GPUs run at higher clock speed and feature a 450W TDP each. The performance testing of the basic and tensor core implementations were limited to the base DGX-2 system. 

Table \ref{weak_scaling} reports the weak scaling measurements obtained running the optimized implementation on both a DGX-2 and a DGX-2H system. The lattice size per GPU has been kept constant at $63.5\times 10^9$ spins (30GB) and we measured the flip/ns rate for $128$ update steps. As expected, the scaling is perfectly linear up to 16 GPUs. This is due to the simple access pattern of remote memory and to the high throughput of the NVLink communications. The performance figures from the table are plotted on the adjacent graph together with the numbers reported in \cite{TPU2019}. Due to memory constraints we couldn't run a test case of size similar to the largest one reported in \cite{TPU2019}.
%so we limited our plot to size $10^{12}$. It should be noted, however, that the TPUv3 code %scaled linearly up to size $3.4\times 10^{12}$, where it achieves $5853$ flip/ns.

Table \ref{strong_scaling} reports the strong scaling measurements of the optimized implementation obtained
on the two DGX systems. The total lattice size has been kept constant at $63.5\times 10^9$ spins (30GB) and it has been partitioned in as many horizontal slabs of the same size as the number of GPUs used in the measurements. As for the weak scaling case, we measured the flip/ns rate for $128$ update steps. Since with any number of GPUs the transfers of the top and of the bottom boundaries is negligible with respect to the processing of the bulk, the scaling is linear up to $16$ GPUs. The plot near the table shows a comparison of the performance figures from the two DGX systems.

% CAN ANYONE MANAGE TO MAKE THIS APPEAR IN THE SAME PAGE AS TABLE 2?
\addtocounter{footnote}{-1}
\stepcounter{footnote}\footnotetext{In \cite{TPU2019} it was incorrectly reported as $0.614$, $1000$ times smaller than the actual value.}

For completeness, we also report the weak and scaling measurements obtained on a DGX-2 system for the basic implementation in Python using MPI and CUDA IPC and the tensor core implementation using unified memory in Table \ref{weak_scaling_2}. Similar to the optimized implementation, both of these implementations achieve very good scaling efficiency. Notably, the tensor core implementation with unified memory achieves better scaling efficiency than the basic implementation using MPI and CUDA IPC.

\begin{table}[ht]
    \begin{minipage}{0.5\linewidth}
    \centering
        \footnotesize
        \begin{tabular}{ccrr}
                      &                                           & \multicolumn{2}{c}{flips/ns} \\ 
             \cmidrule(lr){3-4}
             GPUs     & lattice size                              &    DGX-2 &  DGX-2H           \\ \hline
              1       & $(123\times 2048)\times (123\times 2048)$ &   417.53 &  459.16           \\
              2       & $(246\times 2048)\times (123\times 2048)$ &   828.21 &  916.40           \\
              4       & $(246\times 2048)\times (246\times 2048)$ &  1619.81 & 1831.73           \\
              8       & $(492\times 2048)\times (246\times 2048)$ &  3231.89 & 3661.47           \\
             16       & $(492\times 2048)\times (492\times 2048)$ &  6441.68 & 7381.30           \\ \hline\hline
        \end{tabular}
        \begin{tabular}{ccr}
          TPUv3 cores & lattice size                              &  flip/ns \\ \hline
  $1\times 1\times 2$ &         $(986\times 128)^2$               &    22.89 \\
  $2\times 2\times 2$ &        $(1792\times 128)^2$               &    91.52 \\
  $4\times 4\times 2$ &        $(3584\times 128)^2$               &   366.01 \\
  $8\times 8\times 2$ &        $(7168\times 128)^2$               &  1463.51 \\
$16\times 16\times 2$ &       $(14336\times 128)^2$               &  5853.04 \\\hline
        \end{tabular}
    \end{minipage}
    \hspace{0.1in}
    %\hfill
    \begin{minipage}{0.5\linewidth}
        \centering
        \includegraphics[scale=0.33]{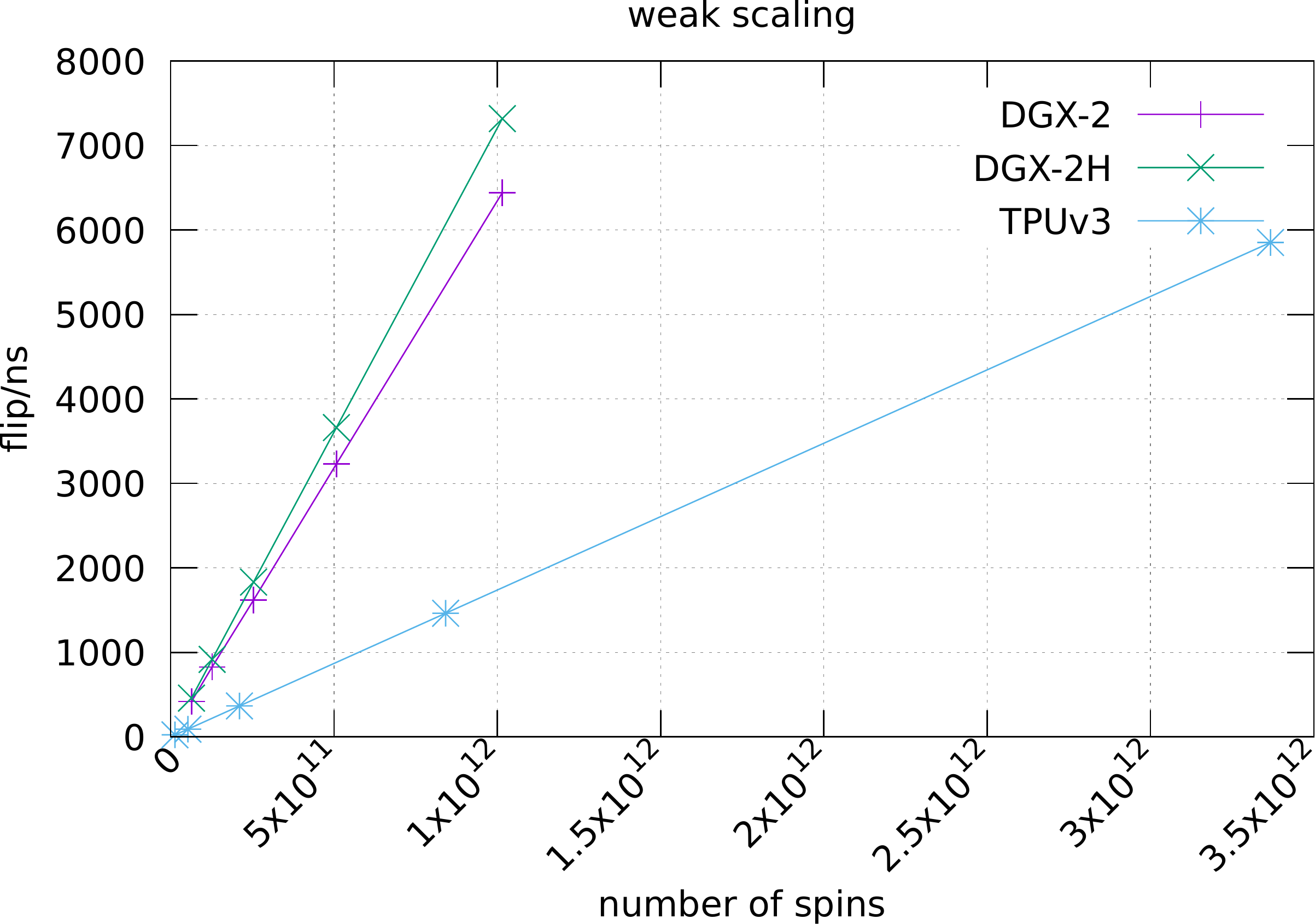}
    \end{minipage}
    \vspace{.1in}
    \caption{Weak scaling of the optimized multi-spin code measured using up to 16 V100-SXM GPUs in a DGX-2 and DGX-2H system, keeping the number of spins per GPU fixed at $(123\times 2048)^2$ (top table). For comparison purposes, the lower table contains the weak scaling measurements reported in \cite{TPU2019} on a multi-TPU system. On the right, are plotted the scaling figures for the DGX systems, and for the TPU system. A single DGX-2 outperforms 64 TPU units (256 chips, 512 cores). Dividing flips/ns by number of cores, the flips per nanosecond per TPUv3 core is roughly 11.43, compared to 417.53 per GPU — more than a $30\times$ speedup.} \label{weak_scaling}
\end{table}

\begin{table}[ht]
    \begin{minipage}{0.50\linewidth}
    \centering
%        \begin{tabular}{ccrr}
%        no. of GPUs & lattice size         & flip/ns  & efficiency \\ \hline
%            1       &                      &   417.57 & 1.000 \\
%            2       &                      &   830.29 & 0.994 \\
%            4       & $(123\times 2048)^2$ &  1629.32 & 0.975\\
%            8       &                      &  3252.68 & 0.973\\ 
%           16       &                      &  6474.16 & 0.969\\ \hline
%        \end{tabular}
        \begin{tabular}{ccrr}
                    &                      & \multicolumn{2}{c}{flips/ns} \\
                    \cmidrule(lr){3-4}
        no. of GPUs &     lattice size     &   DGX-2 &  DGX-2H \\ \hline
            1       &                      &  417.57 &  453.56 \\
            2       &                      &  830.29 &  925.99 \\
            4       & $(123\times 2048)^2$ & 1629.32 & 1848.44 \\
            8       &                      & 3252.68 & 3682.90 \\ 
           16       &                      & 6474.16 & 7292.19 \\ \hline
        \end{tabular}
    \end{minipage}
    \hspace{0.1in}
    %\hfill
    \begin{minipage}{0.5\linewidth}
        \centering
        \includegraphics[scale=0.33]{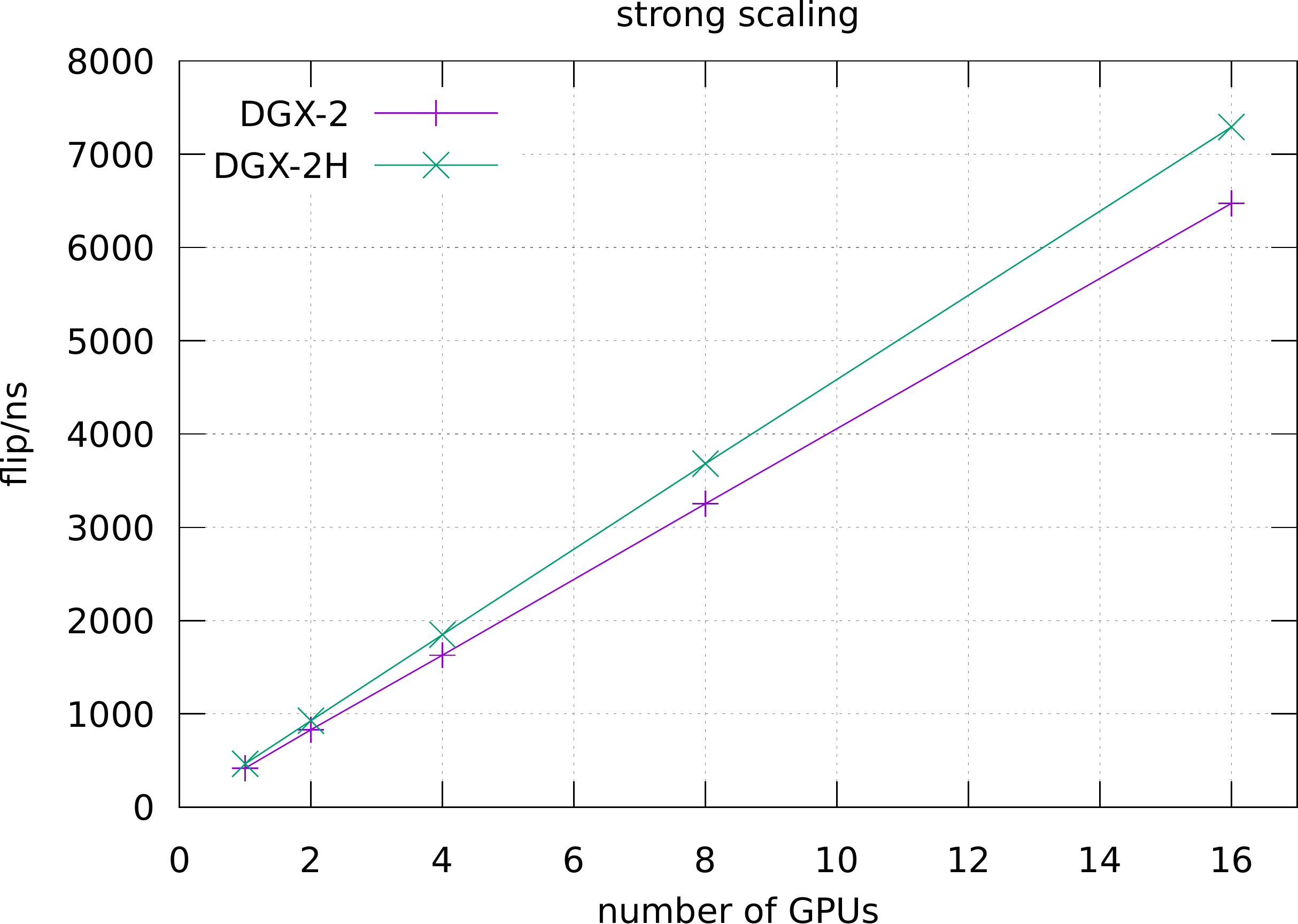}
    \end{minipage}
    \vspace{.1in}
    \caption{Strong scaling of the optimized multi-spin code measured using both a DGX-2 and DGX-2H system with a lattice of fixed size equal to $(123\times 2048)^2$. On the right, it is shown a plot of the two scaling measurements.}
    \label{strong_scaling}
\end{table}

\begin{table}[ht]
    \centering
    \begin{tabular}{ccrr}
        no. of GPUs & lattice size & Basic (Python) & Tensor Core \\
        \hline
        1  & $(640\times128) \times (640\times128)$   & 43.488  & 38.747  \\
        2  & $(1280\times128) \times (640\times128)$  & 82.447  & 77.492  \\
        4  & $(1280\times128) \times (1280\times128)$ & 164.352 & 154.980 \\
        8  & $(2560\times128) \times (1280\times128)$ & 327.136 & 309.918 \\
        16 & $(2560\times128) \times (2560\times128)$ & 648.254 & 619.520 \\ \hline
        1  &                                          & 43.481  & 38.752  \\
        2  &                                          & 83.146  & 78.104  \\
        4  & $(640\times128) \times (640\times128)$   & 165.793 & 156.676 \\
        8  &                                          & 330.258 & 313.077 \\
        16 &                                          & 650.543 & 602.083 \\ \hline
    \end{tabular}
    \vspace{.1in}
    \caption{Weak (top) and strong (bottom) scaling of the basic and tensor core implementations measured using up to 16 V100-SXM GPUs in a DGX-2 system. Results reported in flips/ns.}\label{weak_scaling_2}
\end{table}

\subsection{Validation}

\begin{figure*}[ht]
\begin{center}
        \includegraphics[scale=0.65]{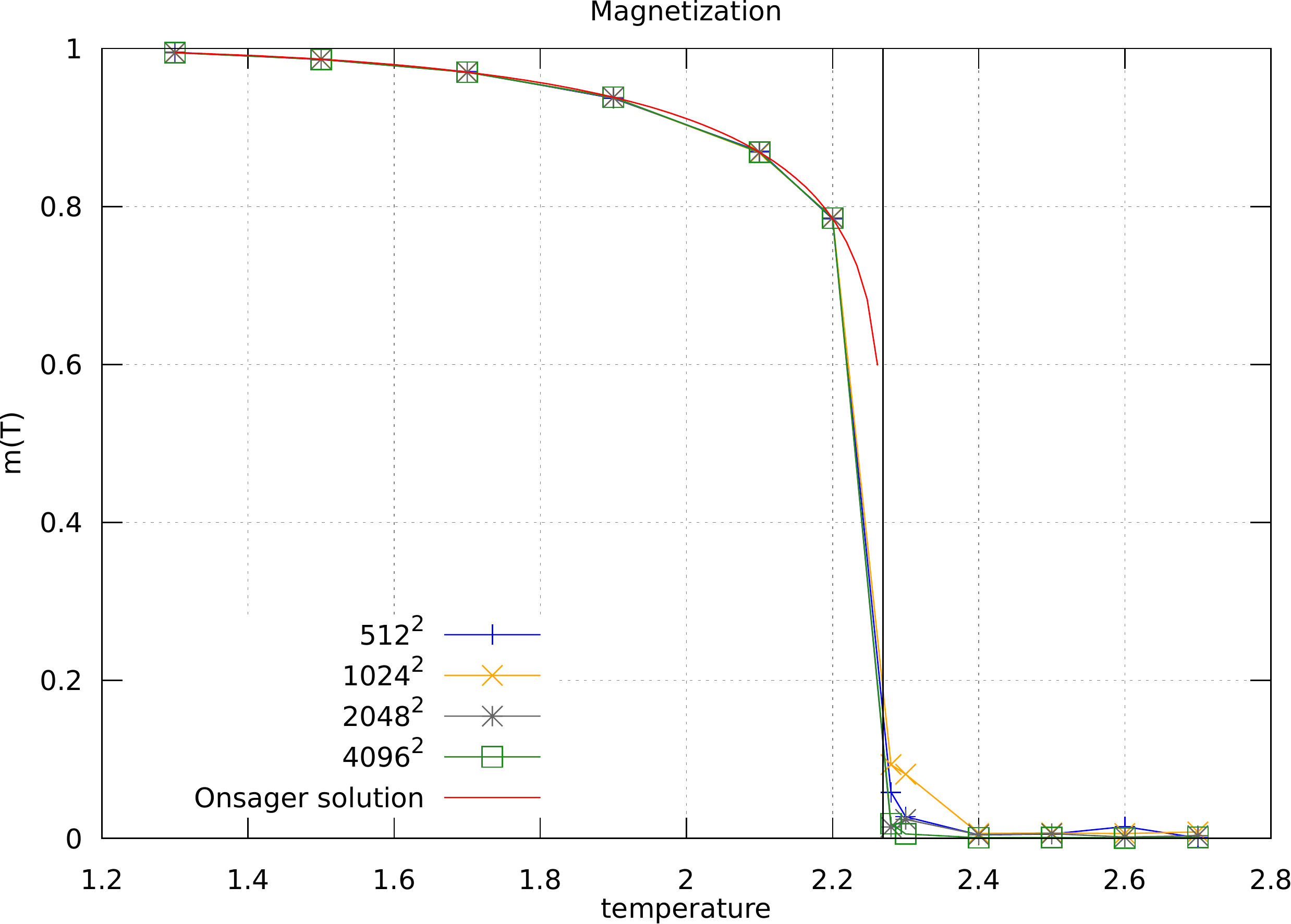}
\end{center}
        \caption{Steady state magnetization measures obtained with the multi-spin code for lattice sizes $512^2$, $1024^2$, $2048^2$, and $4096^2$. The solid vertical line marks the critical temperature value $T_c=2.269185$.
        %(Need vertical line for Tc and plot of the formula)
        }
        \label{fig:magn}
\end{figure*}

\begin{figure*}[t]
\begin{center}
        \includegraphics[scale=0.65]{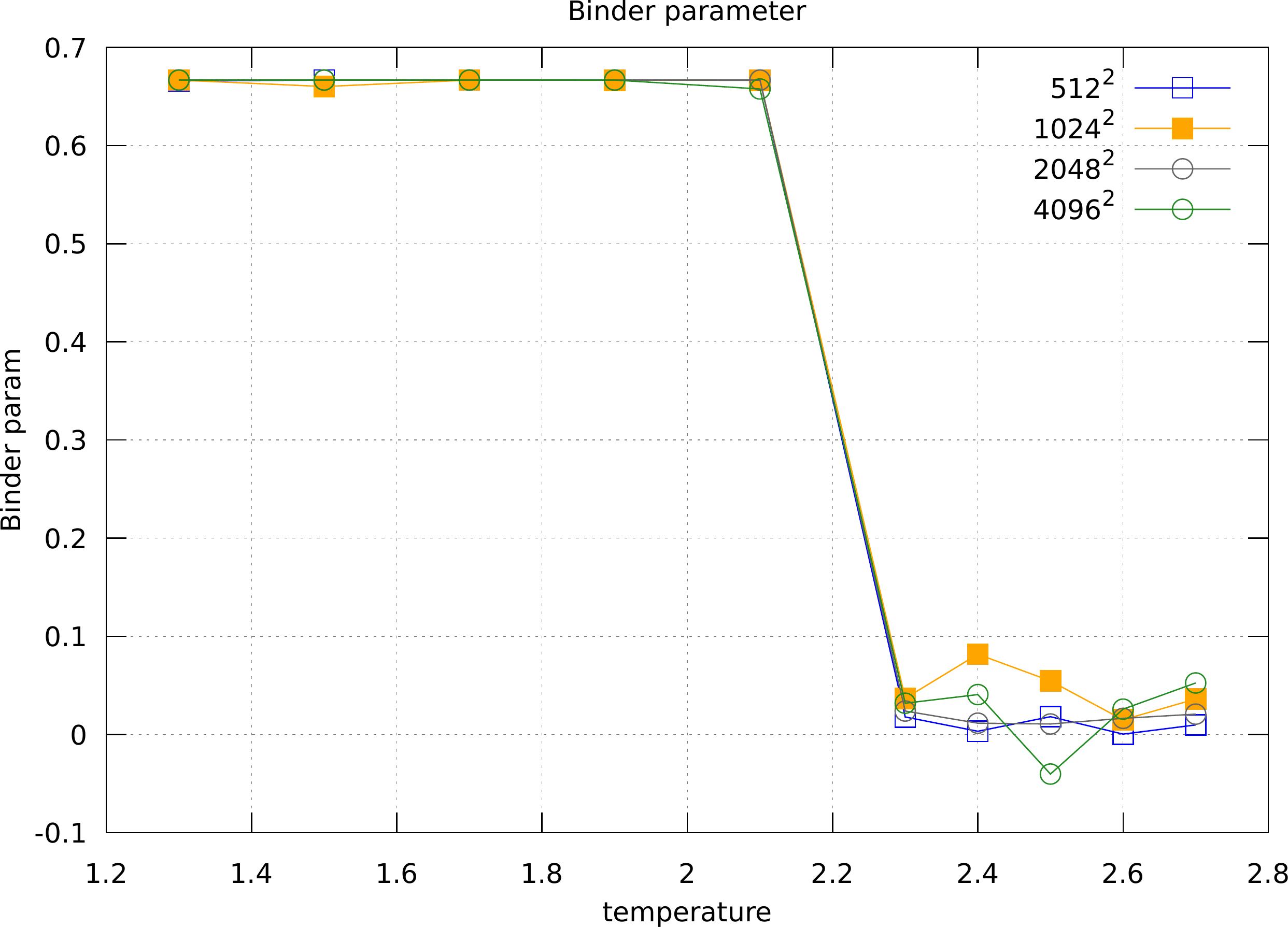}
\end{center}
        \caption{Binder parameter obtained with the multi-spin code for lattice sizes $512^2$, $1024^2$, $2048^2$, and $4096^2$ running for, respectively, 16M, 64M, 256M, and 1B update steps.}
        \label{fig:binder}
\end{figure*}

As mentioned in Section \ref{sec:intro}, there is an analytical solution for the 2D Ising model in the limit of infinite volume that shows the presence of an order-disorder phase transition such that the magnetization is $0$ for $T>T_c$ whereas it is equal to:
\begin{equation}(1-[\sinh{(2J/T)}]^{-4})^{\frac{1}{8}}\label{eqn:magn}
\end{equation} for $T<T_c$. 
The condition for determining the critical temperature $T_c$ at which the phase transition occurs is $(\tanh{(2J/T_c)})^2=1$, corresponding to $T_c=2.269185J$. To verify the accuracy of our optimized code for the simulation of the 2D Ising model, we carried out a set of tests for different temperatures and lattice sizes comparing the results of simulated magnetization with the corresponding (\textit{i.e.,} for the same temperature) values of Equation \ref{eqn:magn}. Figure \ref{fig:magn} shows the simulated magnetization with respect to the temperatures along with the values of the analytical expression \ref{eqn:magn}. Another interesting quantity is the so-called Binder parameter or Binder cumulant \cite{Binder} corresponding to the kurtosis of the order parameter (\textit{i.e.,} the magnetization). The Binder parameter $U_L$ can be computed as $$U_L=1-\frac{<m^4>_L}{(<m^2>)^2_L}$$
This parameter makes it possible to accurately determine phase transition points in numerical simulations of various spin models. The phase transition point may be identified comparing the behavior of $U$ as a function of the temperature for different values of the system size $L$. The critical temperature corresponds to the point where the curves for different values of $L$ cross. Our measures of the Binder parameter are reported in Figure \ref{fig:binder} for the same lattice sizes used for the magnetization. As expected they show a phase transition occurring at the critical temperature $T_c$. Although these measures confirm the accuracy of the simulations using our implementation, we found other interesting effects on large systems ({\em e.g.,} for $L > 1024$) that deserve an in-depth analysis. In particular we observed that in some runs, the time (measured as number of lattice sweeps of the Metropolis algorithm) required to reach the \textit{steady} state is much larger than expected (far from the critical temperature, it should be $\sim L^2$). In those cases, spins tend to organize into bands (horizontal, vertical or even diagonal), remaining in those meta-stable states for very long times. We plan to study that phenomenon in a forthcoming paper.

\section{Conclusions}
We have presented several implementations of the 2D Ising model using a variety of approaches. All the implementations are accurate, fast and scale very well to multi-GPU configurations. We discussed some of the limitations of the basic and tensor core approaches outlined, however, found that all outperformed results on TPU systems \cite{TPU2019}. Our optimized implementation achieves state-of-the-art performance and compares favorably to custom FPGA solutions \cite{FPGA}.  

These codes can be easily extended to simulate other models for which
there are no analytical solutions, for instance a 2D Ising spin glass model, and can also be used to study the dynamics of the classic ({\em i.e.,} ferromagnetic) Ising model simulated with the Metropolis algorithm.

The different versions of the code are available at http://github.com/NVIDIA/ising-gpu.

\section*{Acknowledgements}
We would like to thank Profs. Giorgio Parisi, Enzo Marinari and Federico Ricci-Tersenghi from the University of Rome "Sapienza" for useful discussions about the convergence properties of the Ising model on large lattices.

\bibliographystyle{unsrt}  


\begin{thebibliography}{1}

\bibitem{Ising}
Ising, E. 
\newblock Contribution to the
Theory of Ferromagnetism.
\newblock Zeitschrift für Physik,
vol. XXXI, 1925.

\bibitem{Metropolis}
 Metropolis, N. and Rosenbluth, A.W. and Rosenbluth, M.N. and Teller, A.H. and Teller, E.
 \newblock Equation of State Calculations by Fast Computing Machines.
 \newblock Journal of Chemical Physics. 21 (6): 1087–1092, 1953.

\bibitem{Wolff1989}
Wolff, U. 
\newblock Collective Monte Carlo Updating for Spin Systems.
\newblock Phys. Rev. Lett. 62, 361, 1989.

\bibitem{Bernaschi2012}
Bernaschi, M. and Fatica, M. and Parisi, G. and Parisi, L.
\newblock Multi-GPU codes for spin systems simulations.
\newblock Computer Physics Communications, Vol. 183, 2012.

\bibitem{Onsager}
Onsager, L, 
\newblock Crystal statistics. I. A two-dimensional model with an order-disorder transition.
\newblock Physical Review, Series II, 65 (3–4): 117–149, 1944.

\bibitem{Multispin}Jacobs, L, Rebbi, C.
\newblock Multi-spin coding: a very efficient technique for Monte Carlo simulations of spin systems.
\newblock Journal of Computational Physics 41(1), 203–210, 1981.

\bibitem{TPU2019}
Kun Yang, Yi-Fan Chen, Georgios Roumpos, Chris Colby, John Anderson.
\newblock High Performance Monte Carlo Simulation of Ising Model on TPU Clusters.
\newblock {\em arXiv preprint:1903.11714v2}, 2019.

\bibitem{FPGA}
F. Ortega-Zamorano , M. A. Montemurro, S. A. Cannas, J. M. Jerez, L. Franco.
\newblock FPGA Hardware Acceleration of Monte Carlo Simulations for the Ising Model.
\newblock IEEE Transactions on Parallel and Distributed Systems, 2016, Vol. 27 Issue 9, Pages:2618-2627

\bibitem{Numba}
S. K. Lam, A. Pitrou, Antoine and S. Seibert.
\newblock Numba: A LLVM-based Python JIT Compiler.
\newblock Proceedings of the Second Workshop on the LLVM Compiler Infrastructure in HPC, p.1-6, November 15-15, 2015, Austin, Texas.

\bibitem{CuPy}
R. Okuta, Y. Unno, D. Nishino, S. Hido, and C. Loomis.
\newblock CuPy: A NumPy-Compatible Library for NVIDIA GPU Calculations.
\newblock Proceedings of Workshop on Machine Learning Systems (LearningSys) in The Thirty-first Annual Conference on Neural Information Processing Systems, 2017. 

\bibitem{cuRAND} cuRAND Library, {\em http://docs.nvidia.com/cuda/curand}

\bibitem{cuBLAS} cuBLAS Library, {\em http://docs.nvidia.com/cuda/cublas}

\bibitem{DGX2} NVIDIA DGX-2, {\em https://www.nvidia.com/en-us/data-center/dgx-2/}

\bibitem{Binder}Binder K, Phys.Rev. Lett. 47, 693, 1981.




\end{thebibliography}

\end{document}